\documentclass[twocolumn,amsmath,prl]{revtex4}

\usepackage{dcolumn}
\usepackage{bm}
\usepackage[T1]{fontenc}
\usepackage{amsmath}
\usepackage{graphicx}

\usepackage{hyperref}
\hypersetup{colorlinks=true, linkcolor=blue, citecolor=red, urlcolor=magenta, pdftitle={AO epitax}, pdfauthor={Eric Bousquet}}

\begin{document}
\title{Strain-induced ferroelectricity in simple rocksalt binary oxides}

\author{Eric Bousquet$^1$, Nicola Spaldin$^2$ and Philippe Ghosez$^{1*}$}
\affiliation{$^1$Physique Th\'eorique des Mat\'eriaux, Universit\'e de Li\`ege, B-4000 Sart Tilman, Belgium 
\\
$^2$Materials Department, University of California, Santa Barbara, CA 93106, USA
}

\begin{abstract}
 The alkaline earth binary oxides adopt a simple rocksalt structure and form an important family of compounds because of their large presence in the earth's mantle~\cite{karki1999} and their potential use in microelectronic devices~\cite{hubbard1996}. 
In comparison to the class of multifunctional ferroelectric perovskite oxides \cite{rabe2007}, however, their practical applications remain limited and the emergence of ferroelectricity and related functional properties in simple binary oxides seems so unlikely that it was never previously considered. 
Here, we show using first-principles density functional calculations that ferroelectricity can be easily induced in simple alkaline earth binary oxides such as barium oxide (BaO) using appropriate epitaxial strains. 
Going beyond the fundamental discovery, we show that the functional properties (polarization, dielectric constant and piezoelectric response) of such strained binary oxides are comparable in magnitude to those of typical ferroelectric perovskite oxides, so making them of direct interest for applications. 
Finally, we show that magnetic binary oxides such as EuO, with the same rocksalt structure, behave similarly to the alkaline earth oxides, suggesting a route to new multiferroics combining ferroelectric and magnetic properties.
\end{abstract}

\maketitle

The alkaline earth oxides -- MgO, CaO, SrO and BaO -- have been intensively studied~\cite{oganov2003} and their electronic structures, elasticities, thermal properties and equations of state are well established \cite{schutt1994,posternak1997}.  In view of their relatively wide band gaps and their compatibility with silicon, they recently aroused some interest as possible gate dielectrics to replace SiO$_2$ in MOSFET devices~\cite{hubbard1996}.  Also, they have been shown to play a key role as a buffer layer in the epitaxial growth of multifunctional perovskite oxides directly on silicon~\cite{mckee2001}. As a result, many recent works have been devoted to the study of AO oxide thin films on silicon substrates, with a special emphasis on their lattice mismatch, coherence and band offsets with Si~\cite{mckee2001,forst2003,hubbard1996}. 
Surprisingly, although many studies of the phase diagrams at high pressure have been performed \cite{karki2003}, the phase diagrams of alkaline earth oxides under epitaxial strain have not been previously reported.  

Ferroelectricity in oxides is usually associated with the family of ABO$_3$ perovskites \cite{rabe2007}. The ferroelectric phase transition in this class of materials is described in terms of a displacive transition from a high-symmetry paraelectric phase by condensation of a polar transverse-optical (TO) soft phonon mode below the critical temperature. Experimentally, this is characterized by a decrease in frequency of the TO mode as the ferroelectric transition temperature is approached from above; computationally, the frequency of the TO mode is found to be imaginary in the paraelectric phase, consistent with the absence of a restoring force for ionic displacements \cite{zhong1994b}. As first proposed by Cochran \cite{cochran1960}, the ferroelectric instability can be explained from the compensation of short-range forces favoring the undistorted paraelectric structure by long-range Coulomb interactions favoring the ferroelectric phase. This was confirmed at the first-principles level \cite{ghosez1996}, pointing out that the unusually large destabilizing Coulomb interaction yielding the instability in this class of compounds is linked to giant anomalous Born effective charges, $Z^*$ (2.77 e, 7.25 e and -5.71 e for Ba, Ti and O in BaTiO$_3$  instead of nominal charges of 2.0 e, 4.0 e and -2.0 e), themselves produced by the strong sensitivity of O 2p -- metal d hybridizations to atomic displacements \cite{ghosez1998}. The tendency of materials to ferroelectric instability was also recently reformulated in the framework of vibronic coupling theory \cite{hill2000,rondinelli2009}. There again, the rearrangement of electrons through covalent bond formation when atoms are displaced, typically resulting in anomalous $Z^*$,  appears as an important feature to yield ferroelectricity. 

Alkaline earth oxides exhibit anomalous $Z^*$, related to O 2p -- metal d hybridizations through a mechanism similar to ABO$_3$ compounds \cite{posternak1997}: We obtain values of 2.81 e, 2.49 e and 2.39 e respectively for BaO, SrO and CaO (the nominal charge is 2.0 e in all cases). Nevertheless, tendency to ferroelectricty has never been reported or studied in alkaline earth oxides
. 
 The ferroelectric instablity is known to be strongly sensitive to strain and, recently, it was shown that strain engineering can be used to induce ferroelectricity in otherwise non-ferroelectric ABO$_3$ compounds \cite{haeni2004,bhattacharjee2009}. Here we explore whether the same strategy can be applied to AO compounds, by studying the properties of different binary oxides under epitaxial strain using density functional theory calculations (see Methods section). 

We begin by calculating the TO phonon frequencies of the alkaline earth oxides as a function of epitaxial strain to search for the onset of a softening or instability.
In their non-distorded cubic rocksalt structure, these oxides have a three-fold degenerate TO mode at the $\Gamma$ point.
Under biaxial epitaxial strain, as can be practically achieved in thin films on a substrate, the symmetry is reduced to tetragonal (space group \textit{I4/mmm}, No. 139), inducing a splitting of the cubic TO mode into a single A$_{2u}$ mode polarized along the [001] tetragonal  axis, and a two-fold degenerate E$_u$ mode polarized either along the [100] or [010] axis.

We first consider the case of BaO.
Figure~\ref{fig_combo}.a shows the evolution of the A$_{2u}$ and E$_u$ frequencies with respect to a wide range of epitaxial strain, $\eta=(a_s-a_0)/a_0$ where $a_0$ is the relaxed lattice constant of BaO and $a_s$ is the lattice constant of the substrate. 
We clearly see that the TO mode frequencies are strongly strain-dependent. 
For compressive epitaxial strains ($\eta < 0$), the E$_u$ frequency increases smoothly whereas the A$_{2u}$ frequency decreases strongly until it becomes imaginary at a critical epitaxial strain $\eta_c^z\simeq$-1.3\% (left vertical line of Figure~\ref{fig_combo}.a). 
A similar  behavior is observed for tensile epitaxial strains ($\eta > 0$), but here it is the E$_u$ frequency that becomes imaginary at $\eta_c^{xy}\simeq$1.8\% (right vertical line of Figure~\ref{fig_combo}.a). The existence of a polar TO mode with an imaginary frequency is the fingerprint of ferroelectricity:  The condensation of either the A$_{2u}$ or E$_u$ unstable mode in BaO will induce a spontaneous polarization and yield a ferroelectric ground-state.

\begin{widetext}
\begin{center}
\begin{figure}[htbp!]
\begin{minipage}[t]{1.0\textwidth}
\includegraphics[width=16cm, height=7.4422857cm, angle=0]{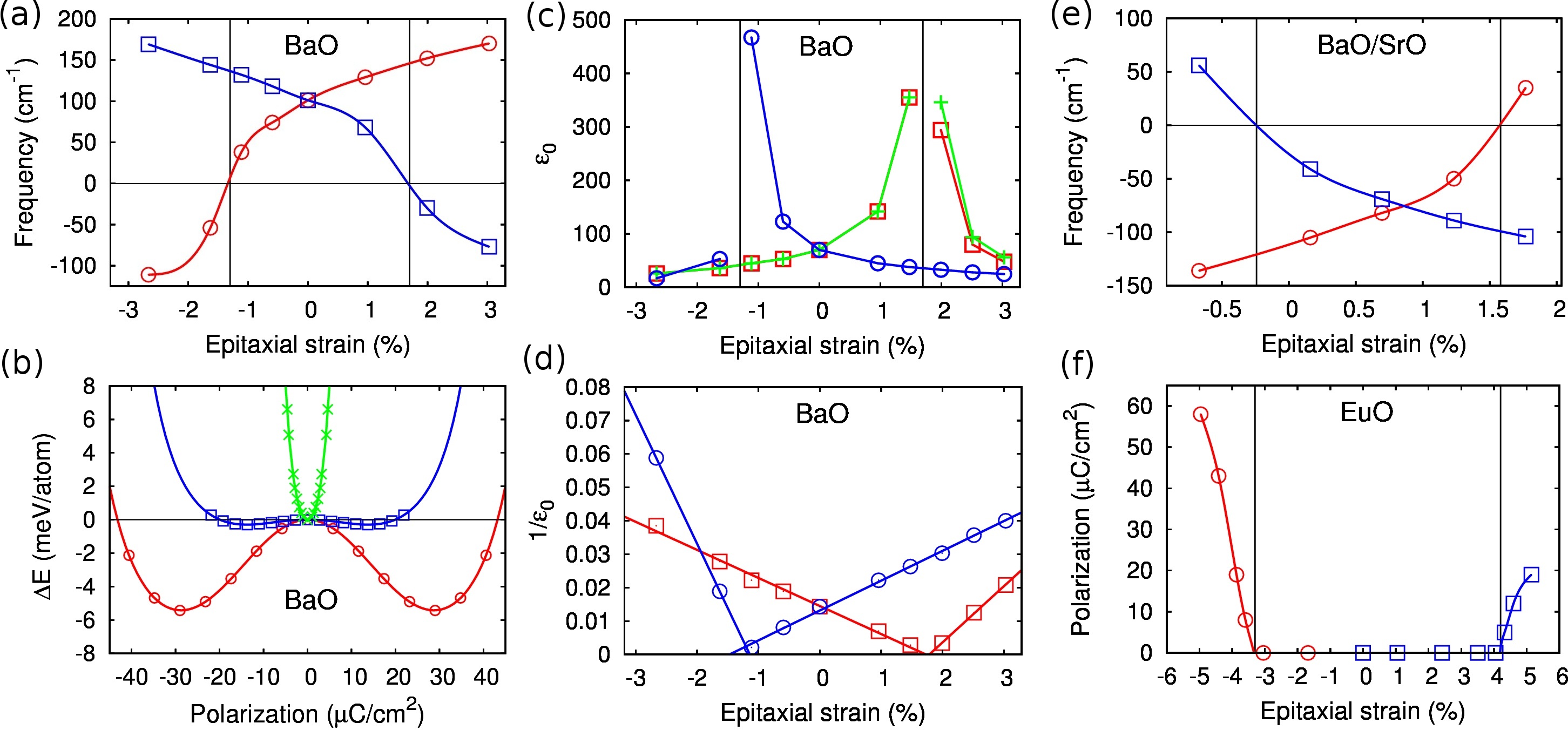} 
\caption{(a) Frequencies of the A$_{2u}$ (red circles) and E$_u$ (blue squares) modes of BaO with respect to the epitaxial strain.
Negative numbers correspond to imaginary frequencies.
(b) Energy as a function of polarization for epitaxial BaO. Green crosses, blue squares and red circles correspond to an epitaxial strain of respectively -0.6\%, -1.63\% and -2.67\%.
(c) BaO static dielectric constant $\varepsilon_0$ versus the epitaxial strain ($\varepsilon_0^{xx}$ red squares, $\varepsilon_0^{yy}$ green crosses, $\varepsilon_0^{zz}$ blue circles) and (d) 1/$\varepsilon_0$ versus the epitaxial strain (1/$\varepsilon_0^{xx}$ red squares, 1/$\varepsilon_0^{zz}$ blue circles).
(e) Frequencies of the A$_{2u}$ (red circles) and E$_u$ (blue squares) modes of BaO/SrO superlattice with respect to the epitaxial strain.
Negative numbers correspond to imaginary frequencies.
(f) Amplitude of the spontaneous polarization of EuO with respect to the epitaxial strain.
Red circles and blue squares are respectively for out-of-plane and in-plane orientation of the polarization.
The vertical lines of graphs a, c, d, e and f give the position of the critical strains $\eta_c^z$ and $\eta_c^{xy}$.}
\label{fig_combo}
\end{minipage}
\end{figure}
\end{center}
\end{widetext}

To determine the ground-state under high epitaxial strains, we next performed full atomic relaxations, intialized by condensing the unstable polar modes, for each mistfit strain exceeding the critical value.
For compressive epitaxial strain, the strongest instability corresponds to polar distortions along the [001] direction (A$_{2u}$). 
The condensation of this instability brings the system to a new ground-state, hereafter called the c-phase, with a non-zero polarization along the [001] direction (P$_z$).
The gain of energy with respect to the paraelectric reference ($\Delta$E), the polarization (P) and the tetragonality $c/a$ are reported in Table~\ref{dataEpitax}.
At an epitaxial strain of -2.66\%, these numbers are huge for an otherwise classical dielectric compound and comparable in amplitude to a typical ferroelectric such as BaTiO$_3$ (P$\simeq$34 $\mu$C.cm$^{-1}$, $\Delta$E$\simeq$3.8 meV/atom).
Figure~\ref{fig_combo}.b shows the evolution of the energy with the polarization resulting from the atomic relaxation at an epitaxial strain of -2.63\% (red circles curve).
As imposed by symmetry, the variation of energy is the same when the polarization is oriented up or down, so that the energy displays the double-well shape characteristic of a ferroelectric compound. 
For comparison, an analogous plot is shown at an epitaxial strain of -0.6\% (green crosses in Figure~\ref{fig_combo}.b) where no polar instability is present. 
Here a single-well shape is obtained, indicating that there is no energy gain when the compound is polarized. 

The absence of any other phonon instability in the polar phase, as explicitly checked from the computation of the full phonon dispersion curves, confirms that the ground-state indeed corresponds to the ferroelectric c-phase when $\eta> \eta_c^z$. 
As in the case of the ferroelectric perovskites, the tetragonality is strongly modified by the polarization; 
we find that the polarization increases the tetragonality ($c/a$=$(c/a)_0$+$\alpha P^2$) of the structure with respect to the paraelectric reference and the change of $c/a$ is continuous at $\eta_c^z$.

\begin{table}[htbp!]
\begin{center}
\begin{tabular}{r@{.}l ccccc}
\hline
\hline
\multicolumn{2}{c}{Misfit-} & phase & $\vert$P$\vert$   & $\Delta$E & ($c/a$)$_0$  & $c/a$    \\
\multicolumn{2}{c}{strain}  &       & ($\mu$C.cm$^{-2}$)& (meV/atom)     &            &        \\
\hline
\ \ -2&66\%       &  c    &  29             &  5.43     &  1.045     & 1.080    \\ 
\ \ -1&63\%       &  c    &  14             &  0.30     &  1.027     & 1.034    \\
\ \  0&00\%       &  -    &  0              &  0.00     &  1.000     & 1.000    \\
\ \ +1&99\%       &  a    &  8              &  0.05     &  0.969     & 0.969    \\
\ \ +2&51\%       &  a    &  15             &  0.13     &  0.962     & 0.959    \\ 
\ \ +3&03\%       &  a    &  21             &  1.88     &  0.954     & 0.950    \\
\hline
\hline
\end{tabular}
\caption{Amplitude of the spontaneous polarization $\vert$P$\vert$, gain of energy $\Delta$E with respect to the paraelectric reference, ratio between the out-of-plane and the in-plane cell parameter (Ba-O distances) of the paraelectric reference $(c/a)_0$ and of the ground-state $c/a$ for five epitaxial strains where the c-phase or the aa-phase are stable.}
\label{dataEpitax}
\end{center}
\end{table}

Similarly, under tensile epitaxial strains, the condensation of the unstable E$_u$ mode for $\eta>\eta_c^{xy}$ yields a ferroelectric ground-state with an in-plane polarization hereafer called the a-phase.
The value of the spontaneous polarization, depth of the double well and tetragonality are reported for the a-phase in Table~\ref{dataEpitax}.
For equivalent strain amplitude, the spontaneous polarization and the depth of the double well are smaller than the corresponding c-phase values, but are still comparable to typical perovskite ferroelectrics. 
The modification of $c/a$ with respect to the paraelectric $(c/a)_0$ is almost negligible with a tendency to decrease when the epitaxial strain or $P$ increases ($\alpha<$0).

Next, we evaluate the effect of epitaxial strain on the functional properties such as the dielectric permittivity and piezoelectric response. 
Figure~\ref{fig_combo}.c shows the evolution of the three components $\varepsilon_0^{xx}$, $\varepsilon_0^{yy}$ and $\varepsilon_0^{zz}$ of the static dielectric tensor with the epitaxial strain.
For $\eta$=0\%, the dielectric tensor is isotropic since the structure is cubic.
For positive and negative epitaxial strain, $\varepsilon_0^{zz}$ becomes different from $\varepsilon_0^{xx}$ and $\varepsilon_0^{yy}$.
The value of $\varepsilon_0^{zz}$ (resp. $\varepsilon_0^{xx}$ and $\varepsilon_0^{yy}$) diverges around $\eta_c^z$ (resp. $\eta_c^{xy}$).
This divergent behavior of $\varepsilon_0$, exemplified from the linear evolution of 1/$\varepsilon_0$ in Figure~\ref{fig_combo}.d, is a usual feature of a displacive ferroelectric phase transition and is related to the softening of a transverse optic polar mode (Eq.55 of Ref~\cite{gonze1997}).

Since the ground-state is non-centrosymmetric beyond the critical epitaxial strain $\eta_c$, it will also be piezoelectric. 
In Table~\ref{piezo} we report the piezoelectric stress constants e$_{ij}$ of BaO for selected epitaxial strains in the ferroelectric region. 
For negative epitaxial strains, in the c-phase, the dominant coefficient is e$_{33}$. 
This coefficient is large and is comparable to that of ferroelectric perovskites such as PbTiO$_3$ (3.68 C.m$^{-2}$~\cite{saghi-szabo1998}), BaTiO$_3$ (3.64  C.m$^{-2}$~\cite{zgonik1994}) or even the widely used Pb(Zr,Ti)O$_3$ (PZT) alloys (3.4 C.m$^{-2}$~\cite{bellaiche1999}). In the a-phase for tensile epitaxial strains, the largest piezoelectric constants are  e$_{11}$, e$_{12}$, e$_{13}$ and e$_{26}$ .

\begin{table}[htbp!]
\begin{center}
\begin{tabular}{c r@{.}l r@{.}l r@{.}l r@{.}l }
\hline
\hline
                 & -2&67\% & -1&63\% & 1&99\% & 3&03\%  \\
\hline
e$_{11}$         & 0&00                   & 0&00	          &  \textbf{8}&\textbf{13} &	\textbf{3}&\textbf{37}    \\
e$_{31}$=e$_{32}$&-0&96                   & -2&08	          &  0&00                   &	 0&00    \\
e$_{12}$         & 0&00                   & 0&00	          &  \textbf{8}&\textbf{02} &	\textbf{3}&\textbf{42}  \\
e$_{13}$         & 0&00                   & 0&00	          & \textbf{-6}&\textbf{77} &	 \textbf{-2}&\textbf{66}  \\
e$_{33}$         & \textbf{3}&\textbf{19} & \textbf{6}&\textbf{24}&  0&00  &	 0&00    \\
e$_{24}$=e$_{15}$&-0&29                   & -0&17	          &  0&00  &	 0&00    \\
e$_{35}$         & 0&00                   & 0&00	          & -0&05  &	 -0&14    \\
e$_{26}$         & 0&00                   & 0&00	          & \textbf{14}&\textbf{05}  &	\textbf{5}&\textbf{30}  \\
\hline
\hline
\end{tabular}
\caption{Non-zero piezoelectric stress constants e$_{ij}$ (C.m$^{-2}$) of BaO at epitaxial strains of -2.67\%, -1.63\%, 1.99\% and 3.03\%.}
\label{piezo}
\end{center}
\end{table}

Now that we have demonstrated the possibility of inducing ferroelectricity in BaO thin films under epitaxial strain, we  investigate if the phenomenon is specific to BaO or a more general feature shared with other alkaline earth compounds. 
Looking at  CaO and SrO, we find a very similar behavior. 
Compressive (resp. tensile) epitaxial strain makes the A$_{2u}$ (resp. E$_u$) mode softer. However, since at the bulk level, the cubic TO frequencies of SrO and CaO are higher than in BaO, the calculated critical epitaxial strains are also larger : $\eta_c^z\simeq$-5.1\%  (resp. -6.7\%) and $\eta_c^{x,y}\simeq$ 6.1\% (resp. 7.7\%) in SrO (resp. CaO).

Although the larger critical strains make CaO and SrO  less attractive for applications than BaO, the natural tendency to ferroelectricity of the class of alkaline earth compounds suggests tuning of the ferroelectric properties in superlattices made of the repetition of different oxide layers; such behavior was realized recently in ABO$_3$ compounds~\cite{dawber2007, bousquet2008}. 
As a prototypical example, we consider here a BaO/SrO superlattice built by alternating one layer of BaO and one layer of SrO along the [001] direction. 
The lattice mismatch between BaO and SrO is relatively large ($\sim$7.7\%) and will prevent the growth of large period superlattices but should not be problematic in ultra-short period systems. 
The value of the relaxed in-plane cell parameter of the BaO/SrO superlattice in its paraelectric phase is close to the average of the BaO and SrO cell parameters (see Supplementary Information).
At this relaxed cell parameter, BaO will be under compressive strain and SrO under tensile strain. 
Since such epitaxial strains will favor the c-phase in BaO and the a-phase in SrO, we can expect a competition between the two orientations of the polarization in the superlattice.
In the following, all the epitaxial strains reported are calculated with respect to the relaxed in-plane cell parameter of the superlattice in its paraelectric phase.

In Figure~\ref{fig_combo}.e we show the evolution of the A$_{2u}$ and E$_u$ modes with respect to epitaxial strain. 
We find that \textit{both} of them are unstable for a range of epitaxial strains from -0.24\% to 1.58\%; outside this range {\it either} A$_{2u}$ or E$_u$ is unstable.
This strongly contrasts with the case of individual AO compounds, where the A$_{2u}$ and E$_u$ modes are never unstable together, and the regions in which they are respectively unstable are separated by a range of epitaxial strain without any instability.
This feature makes the superlattice even more interesting: Whatever the epitaxial strain, we expect the ground-state to be ferroelectric.

This is confirmed by the structural relaxation which produces the following sequence of ferroelectric ground-state for the BaO/SrO superlattice: (\textit{i}) for $\eta<$0.07\%, the ground-state is ferroelectric with polarization along the out-of-plane direction (c-phase); (\textit{ii}) for 0.07\%$<\eta<$1.15\%, the ground-state corresponds to a monoclinic phase where the polarization has a component in the in-plane direction and one in the out-of-plane direction giving rise to a small relaxation of the angle between the $x$ and $z$ axis (ac-phase); (\textit{iii}) for $\eta>$1.15\%, the ground-state is ferroelectric with polarization along the in-plane direction (a-phase).

Having demonstrated that ferroelectricity can be strain engineered in different alkaline-earth oxides, we now investigate the possibility of inducing ferroelectricity in the ferromagnetic binary oxide EuO which crystallizes in the same rocksalt structure, as a novel route to multiferroism. Indeed, as for the non-magnetic oxides, we find that EuO becomes ferroelectric under epitaxial strain. Importantly, its magnetic state remains ferromagnetic through the ferroelectric region. The calculated critical epitaxial strains for EuO are $\eta_c^z\simeq$-3.3\% and $\eta_c^{x,y}\simeq$ 4.2\%. These critical strains are larger than those of BaO but remain experimentally achievable. In Figure~\ref{fig_combo}.f we show the calculated evolution of the spontaneous polarization with respect to the epitaxial strain. Here, again, the amplitude of the polarization can reach sizable values (60 $\mu$C.cm$^{-2}$ at -5.5\%) with out-of-plane or in-plane orientation for respectively compressive and tensile epitaxial strain. 

We hope that our theoretical findings will motivate experimentalists to confirm our prediction that ferroelectricity  and related dielectric functionalities can be induced in AO  compounds of rocksalt structure using epitaxial strain. Indeed, it should be possible to tune or modify the properties in thin films and superlattices by appropriate choice of substrate lattice constant. 
If the epitaxial strain is bigger than the critical value $\eta_c$, a ferroelectric ground-state with non-zero polarization and piezoelectric response should be observed.  
If the experimental strain is substantial, but smaller than the critical value, the material will be paraelectric, but its TO mode frequency and dielectric constant should be strongly strain-dependent as reported in Figure~\ref{fig_combo}.d.

Strain-induced ferroelectricity in simple rocksalt AO oxides comparable to that in ABO$_3$ oxides is not only an astonishing fundamental finding but also of direct interest for practical applications. 
Although ferroelectric perovskites are considered as potentially interesting high-k gate dielectrics for MOSFET transistors, a significant drawback is their small bandgap which results in unsuitable band offsets with Si. This problem is directly overcome with strained AO compounds.
Interestingly, we predict that BaO films epitaxially grown on Si ($\eta$=-2.0\%) should exhibit a ferroelectric ground-state under appropriate electrical boundary conditions~\cite{junquera2008}.
In the context of multiferroism, the possibility to induce ferroelectricity in simple magnetic oxides is particularly attractive; previous searches have focussed on more complex compounds~\cite{bhattacharjee2009, rondinelli2009}. 
The case of EuO is particularly interesting since together with inducing ferroelectricity, the epitaxial strain is expected to increase the ferromagnetic Curie temperature~\cite{ingle2008},  providing a route to higher temperature multiferroics.
The possibility to further tune the properties in superlattices combining different AO compounds offers tremendously promising and still unexplored possibilities.

\section*{Method}
For the alkaline earth compounds, we performed first-principles calculations within the local-density approximation to density-functional theory. 
We used the plane-wave based  ABINIT~\cite{abinit} software with norm-conserving pseudopotentials~\cite{teter1993} and a high convergence was obtained with a plane-wave energy cutoff of 50 Ha and a grid of 6$\times$6$\times$6 special k-points to sample the Brillouin zone. 
The vibrational properties, Born effective charges, dielectric tensors and piezoelectric constants were calculated using the density functional perturbation theory~\cite{gonze1997}.

For EuO, we performed the calculation with the VASP code~\cite{vasp2} and using the GGA PBE functional~\cite{perdew1996} with the DFT+U~\cite{liechtenstein1995} technique on the $f$ orbitals of Eu (U=7 eV and J=1 eV).
Good convergence was obtained with an energy cutoff of 600 eV and a 6$\times$6$\times$6 grid of kpoints.

Work at Li\`ege University was supported by EMMI and MaCoMuFi European Project and work at UC Santa Barbara was supported by the National Science Foundation.


\begin{thebibliography}{10}
\bibitem{karki1999}
B.~B. Karki, R.~M. Wentzcovitch, S.~de~Gironcoli and S.~Baroni,
  \emph{First-principles determination of elastic anisotropy and wave
  velocities of {MgO} at lower mantle conditions}, Science \textbf{286}, 1705
  (1999).

\bibitem{hubbard1996}
K.~J. Hubbard and D.~G. Schlom, \emph{Thermodynamic stability of binary oxides
  in contact with silicon}, J. Mater. Res. \textbf{11}, 2757 (1996).

\bibitem{rabe2007}
K.~Rabe, C.~H. Ahn and J.-M. Triscone, eds., \emph{Physics of Ferroelectrics, A
  modern Perspective} (Springer, 2007).

\bibitem{oganov2003}
A.~R. Oganov and P.~I. Dorogokupets, \emph{All-electron and pseudopotential
  study of {MgO}: Equation of state, anharmonicity, and stability}, Phys. Rev.
  B \textbf{67}, 224110 (2003).

\bibitem{schutt1994}
O.~Sch{\"u}tt, P.~Pavone, W.~Windl, K.~Karch and D.~Strauch, \emph{Ab initio
  lattice dynamics and charge fluctuations in alkaline-earth oxides}, Phys.
  Rev. B \textbf{50}, 3746 (1994).

\bibitem{posternak1997}
M.~Posternak, A.~Baldereschi, H.~Krakauer and R.~Resta, \emph{Non-nominal value
  of the dynamical effective charge in alkaline-earth oxides}, Phys. Rev. B
  \textbf{55}, R15983 (1997).

\bibitem{mckee2001}
R.~A. McKee, F.~J. Walker and M.~F. Chisholm, \emph{Physical structure and
  inversion charge at a semiconductor interface with a crystalline oxide},
  Science \textbf{293}, 468 (2001).

\bibitem{forst2003}
C.~J. F{\"o}rst, C.~R. Ashman, K.~Schwarz and P.~E. Bl{\"o}chl, \emph{The
  interface between silicon and a high-k oxide}, Nature \textbf{427}, 53
  (2003).

\bibitem{karki2003}
B.~B. Karki and R.~M. Wentzcovitch, \emph{Vibrational and quasiharmonic thermal
  properties of {CaO} under pressure}, Phys. Rev. B \textbf{68}, 224304 (2003).

\bibitem{zhong1994b}
W.~Zhong, R.~D. King-Smith and D.~Vanderbilt, \emph{Giant {LO-TO} splittings in
  perovskite ferroelectrics}, Phys. Rev. Lett. \textbf{72}, 3618 (1994).

\bibitem{cochran1960}
W.~Cochran, \emph{Crystal stability and the theory of ferroelectricity}, Adv.
  Phys. \textbf{9}, 387 (1960).

\bibitem{ghosez1996}
P.~Ghosez, X.~Gonze and J.-P. Michenaud, \emph{Coulomb interaction and
  ferroelectric instability of {BaTiO$_3$}}, Europhys. Lett. \textbf{33}, 713
  (1996).

\bibitem{ghosez1998}
P.~Ghosez, J.-P. Michenaud and X.~Gonze, \emph{Dynamical atomic charges: The
  case of {ABO$_3$} compounds}, Phys. Rev. B \textbf{58}, 6224 (1998).

\bibitem{hill2000}
N.~A. Hill, \emph{Why are there so few magnetic ferroelectrics?}, J.
  Phys. Chem. B \textbf{104}, 6694 (2000).

\bibitem{rondinelli2009}
J.~M. Rondinelli, A.~S. Eidelson and N.~A. Spaldin, \emph{Non-d$^0$ {M}n-driven
  ferroelectricity in antiferromagnetic {BaMnO$_3$}}, Phys. Rev. B
  \textbf{79}, 205119 (2009).

\bibitem{haeni2004}
J.~H. Haeni, P.~Irvin, W.~Chang, R.~Uecker, P.~Reiche, Y.~L. Li1, S.~Choudhury,
  W.~Tian, M.~E. Hawley, B.~Craigo, A.~K. Tagantsev, X.~Q. Pan, S.~K.
  Streiffer, L.~Q. Chen, S.~W. Kirchoefer, J.~Levy and D.~G. Schlom,
  \emph{Room-temperature ferroelectricity in strained {SrTiO$_3$}}, Nature
  \textbf{430}, 758 (2004).

\bibitem{bhattacharjee2009}
S.~Bhattacharjee, E.~Bousquet and P.~Ghosez, \emph{Engineering multiferroism in
  {CaMnO$_3$}}, Phys. Rev. Lett. \textbf{102}, 117602 (2009).

\bibitem{gonze1997}
X.~Gonze and C.~Lee, \emph{Dynamical matrices, {B}orn effective charges,
  dielectric permittivity tensors, and interatomic force constants from
  density-functional perturbation theory}, Phys. Rev. B \textbf{55}, 10355
  (1997).

\bibitem{saghi-szabo1998}
G.~S{\'a}ghi-Szab{\'o}, R.~E. Cohen and H.~Krakauer, \emph{First-principles
  study of piezoelectricity in {PbTiO$_3$}}, Phys. Rev. Lett. \textbf{80}, 4321
  (1998).

\bibitem{zgonik1994}
M.~Zgonik, P.~Bernasconi, M.~Duelli, R.~Schlesser, P.~G{\"u}nter, M.~H.
  Garrett, D.~Rytz, Y.~Zhu and X.~Wu, \emph{Dielectric, elastic, piezoelectric,
  electro-optic, and elasto-optic tensors of {BaTiO$_3$} crystals}, Phys. Rev.
  B \textbf{50}, 5941 (1994).

\bibitem{bellaiche1999}
L.~Bellaiche and D.~Vanderbilt, \emph{Intrinsic piezoelectric response in
  perovskite alloys: {PMN-PT} versus {PZT}}, Phys. Rev. Lett. \textbf{83}, 1347
  (1999).

\bibitem{dawber2007}
M.~Dawber, N.~Stucki, C.~Lichtensteiger, S.~Gariglio, P.~Ghosez and J.-M.
  Triscone, \emph{Tailoring the properties of artificially layered
  ferroelectric superlattices}, Adv. Mater \textbf{19}, 4153 (2007).

\bibitem{bousquet2008}
E.~Bousquet, M.~Dawber, N.~Stucki, C.~Lichtensteiger, P.~Hermet, S.~Gariglio,
  J.-M. Triscone and P.~Ghosez, \emph{Improper ferroelectricity in perovskite
  oxide artificial superlattices}, Nature \textbf{452}, 732 (2008).

\bibitem{junquera2008}
J.~Junquera and P.~Ghosez, \emph{First-principles study of ferroelectric oxide
  epitaxial thin films and superlattices: Role of the mechanical and electrical
  boundary conditions}, J. Comput. Theor. Nanosci. \textbf{5}, 2071 (2008).

\bibitem{ingle2008}
N.~J.~C. Ingle and I.~S. Elfimov, \emph{Influence of epitaxial strain on the
  ferromagnetic semiconductor {EuO}: First-principles calculations}, Phys.
  Rev. B \textbf{77}, 121202 (2008).

\bibitem{teter1993}
M.~Teter, \emph{Additional condition for transferability in pseudopotentials},
  Phys. Rev. B \textbf{48}, 5031 (1993).

\bibitem{abinit}
X.~Gonze, J.-M. Beuken, R.~Caracas, F.~Detraux, M.~Fuchs, G.-M. Rignanese,
  L.~Sindic, M.~Verstraete, G.~Zerah, F.~Jollet, M.~Torrent, A.~Roy, M.~Mikami,
  P.~Ghosez, J.-Y. Raty and D.~Allan, \emph{First-principles computation of
  material properties : the {ABINIT} software project,}, Comput.
  Mat. Sci. \textbf{25}, 478 (2002).

\bibitem{vasp2}
G.~Kresse and D.~Joubert, \emph{From ultrasoft pseudopotentials to the
  projector augmented-wave method}, Phys. Rev. B \textbf{59}, 1758 (1999).

\bibitem{perdew1996}
J.~P. Perdew, K.~Burke and M.~Ernzerhof, \emph{Generalized gradient
  approximation made simple}, Phys. Rev. Lett. \textbf{77}, 3865 (1996).

\bibitem{liechtenstein1995}
A.~I. Liechtenstein, V.~I. Anisimov and J.~Zaanen, \emph{Density-functional
  theory and strong interactions: Orbital ordering in {M}ott-{H}ubbard
  insulators}, Phys. Rev. B \textbf{52}, R5467 (1995).

\end{thebibliography}
\end{document}